\documentclass[twocolumn,american,aps,prl,a4paper]{revtex4-1}
\usepackage{lmodern}
\usepackage[T1]{fontenc}
\usepackage[latin9]{inputenc}
\usepackage{color}
\usepackage{float}
\usepackage{amsmath}
\usepackage{amssymb}
\usepackage{wasysym}
\usepackage{graphicx}
\usepackage{esint}

\makeatletter
\@ifundefined{textcolor}{}
{%
 \definecolor{BLACK}{gray}{0}
 \definecolor{WHITE}{gray}{1}
 \definecolor{RED}{rgb}{1,0,0}
 \definecolor{GREEN}{rgb}{0,1,0}
 \definecolor{BLUE}{rgb}{0,0,1}
 \definecolor{CYAN}{cmyk}{1,0,0,0}
 \definecolor{MAGENTA}{cmyk}{0,1,0,0}
 \definecolor{YELLOW}{cmyk}{0,0,1,0}
}

\newcommand{\KK}{{\kappa}}

\usepackage{babel}

\makeatother

\usepackage{babel}
\begin{document}

\title{Weyl node with random vector potential}
\begin{abstract}
We study Weyl semimetals in the presence of generic disorder, consisting of a random vector potential as well as a random scalar potential. We derive renormalization group flow equations to second order in the disorder strength. These flow equations predict a disorder-induced phase transition between a pseudo-ballistic weak-disorder phase and a diffusive strong-disorder phase for sufficiently strong random scalar potential or for a pure three-component random vector potential. We verify these predictions using a numerical study of the density of states near the Weyl point and of quantum transport properties at the Weyl point. In contrast, for a pure single-component random vector potential the diffusive strong-disorder phase is absent.
\end{abstract}

\author{Bj\"orn Sbierski, Kevin S.C. Decker, Piet W. Brouwer}

\affiliation{Dahlem Center for Complex Quantum Systems and Institut f\"ur Theoretische
Physik, Freie Universit\"at Berlin, D-14195, Berlin, Germany}

\date{\today}

\maketitle

\paragraph*{Introduction.---}

Weyl semimetals are semimetals with topologically protected non-degenerate band touching points. Their most prominent hallmarks include the ``Fermi arc'' surface states and the chiral anomaly \cite{Wan2011,Hosur2013,Burkov2015b}. Experimental evidence for these signatures has been reported for a wide range of material systems \cite{Bernevig2015,Xu2015,Lv2015b,Xu2015d,Zhang2015}. The low-energy physics of electrons in the vicinity of the band touching points, which are referred to as ``Weyl nodes'', is described by the Weyl Hamiltonian
\begin{equation}
H_{0}=\hbar v\boldsymbol{\sigma}\cdot\mathbf{k}-\sigma_{0}\mu,\label{eq:H0}
\end{equation}
where $v$ is the Fermi velocity, $\boldsymbol{\sigma} = (\sigma_x,\sigma_y,\sigma_z)$ the vector of Pauli matrices, $\sigma_{0}$ the $2\times2$ unit matrix, $\mathbf{k}$ the reciprocal-space distance to the Weyl point, and $\mu$ the chemical potential. While for the specific Weyl semimetals that have been realized to date the Fermi surface is typically small, but not point-like --- {\em i.e.}, the chemical potential $\mu$ is small but nonzero ---, a new generation of ``ideal'' Weyl materials promises a Fermi energy pinned exactly to the nodal point $\mu = 0$ \cite{Ruan2016,Ruan2016a}.

However, even for such ``ideal'' materials,
the presence of impurities and other forms of disorder is unavoidable in a realistic sample. We here consider the case that the disorder is sufficiently smooth, so that it does not couple different Weyl nodes and may be described as an effective (matrix-valued) potential $U(\mathbf{r})$ added to the Weyl Hamiltonian (\ref{eq:H0}). The disorder physics at a Weyl node is very rich and well studied for potential disorder $U(\mathbf{r}) \propto \sigma_{0}$. Already in the 1980s Fradkin predicted the existence of a disorder-induced quantum phase transition from a semimetallic to a diffusive metallic phase with increasing disorder strength \cite{Fradkin1986a,Fradkin1986}. With the renewed interest in Weyl materials, recently this paradigm has inspired further theoretical studies(for a review see \cite{Syzranov2016c}), including field theory developments and prediction of transport properties \cite{Sbierski2014a,Syzranov2015a,Pixley2015a,Pixley2015c}, the study of critical exponents \cite{Kobayashi2014,Syzranov2016,Sbierski2015,Louvet2016}, and the investigation of various shapes for effective impurity potentials \cite{Skinner2014,Ominato2015}.

In this paper, we aim at broadening the scope of the above discussion by studying a Weyl node with {\em generic} disorder $U(\mathbf{r})$ with scalar and vector contributions,
\begin{equation}
  U(\mathbf{r}) = U_0(\mathbf{r}) + \sum_{i=x,y,z} U_{i}(\mathbf{r}) \sigma_{i}.\label{eq:U(r)}
\end{equation}
Unlike the scalar component $U_0$, the vector components $U_{x,y,z}$ break the effective time-reversal symmetry $H_0(\mathbf{k}) = \sigma_y H_0(-\mathbf{k})^* \sigma_y$ of the Weyl Hamiltonian (\ref{eq:H0}). This symmetry is, however, an accidental symmetry of the Hamitonian (\ref{eq:H0}), so that we expect that a random vector potential occurs generically in the effective low-energy description of a disordered Weyl node. (Note that generic Weyl points do not occur at high-symmetry points of the Brillouin zone so that time-reversal symmetry, even if present, does not impose constraints on the Hamiltonian for a single Weyl node.) Furthermore, time-reversal symmetry is broken in some recent material realizations of Weyl semimetals, such as the compounds $\mathrm{YbMnBi_{2}}$ or $\mathrm{SrMnSb}$ \cite{Borisenko2015,Liu2015a}.

The occurrence of a diffusive phase for strong potential disorder can be understood from inspection of the Hamiltonian $H_0 + U_0(\mathbf{r}) \sigma_0$ with a slowly varying potential $U_0$: A large enough potential fluctuation can generate a carrier density at the Weyl point by trapping wavepackets on a length scale $\ell$ comparable to the correlation length $\xi$ of the random potential. At the same time, the stability of the semimetallic phase at weak disorder follows essentially from the scaling dimension of the disorder term $U_0(\mathbf{r})$: The potential energy ($\propto \ell^{-3/2}$) available to confine a wavepacket at a length scale $\ell \gg \xi$ decreases faster than its kinetic energy ($\propto \ell^{-1}$). Whereas the scaling considerations at weak disorder directly carry over to the vector case, the argument for a diffusive phase at strong disorder does not: A slowly varying vector potential merely shifts the location of the Weyl point, but has no effect on the density of states or on transport properties. Indeed, we find that vector disorder comes with a richer phase diagram than scalar disorder: For single-component vector disorder ({\em e.g.,} $U(\mathbf{r}) \propto \sigma_x)$ we find no signs of a diffusive phase. On the other hand, for vector disorder with two or three statistically independent components --- {\em i.e.}, for the generic case ---, there is a diffusive phase above a critical disorder strength. In this case, the phenomenology is the same as in the potential disorder case. 

Our findings are independently based on three different approaches: a scaling theory perturbative in the disorder strength, numerical calculations of the density of states for a lattice model, and transport calculations for a single Weyl node. Below, the three approaches will be discussed separately. We remark that the self-consistent Born approximation, which qualitatively (but not quantitatively) describes the effect of potential disorder, is known to fail for vector disorder in the two-dimensional Dirac problem \cite{Fedorenko2012}, which is why we abstain from using it for the three-dimensional Weyl problem. 

\paragraph*{Scaling analysis.---} An understanding of the qualitative features of a Weyl node with generic disorder can be obtained using a momentum shell renormalization group approach. This method was applied to a Weyl node with scalar disorder by Syzranov {\em et al.} \cite{Syzranov2015a}. The generalization to the generic case is straightforward, and we focus on the main ideas and results here. We consider an effective Hamiltonian in which states with energy above a cutoff $|\varepsilon|>\hbar v\Lambda$ have been integrated out. We assume the disorder to be Gaussian with zero mean and with correlation function 
\begin{eqnarray}
\left\langle U_{\mu}(\mathbf{r})U_{\nu}(\mathbf{r}^{\prime})\right\rangle  & = & \delta_{\mu\nu}\KK_{\mu}\frac{\left(\hbar v\right)^{2}}{\Lambda}\delta\left(\mathbf{r}-\mathbf{r}^{\prime}\right), \label{eq:U_correlator}
\end{eqnarray}
where $\mu,\nu=0,x,y,z$ and the $\KK_{\mu} \ge 0$ are the dimensionless disorder strengths. Performing the disorder average and using the replica trick \cite{Altland2006} one arrives at an effective action $S=S_{0}+\sum_{\mu}S_{{\rm dis},\mu}$ with 
\begin{eqnarray*}
S_{0} & = & \sum_{a}\int \frac{\mathrm{d}\mathbf{q}}{(2\pi)^3} \frac{\mathrm{d} \omega}{ 2 \pi} \,\bar{\psi}_{a}(\mathbf{q},\omega)\, (H_0(\mathbf{q}) - i \omega) \psi_{a}(\mathbf{q},\omega),
\end{eqnarray*}
where $\psi_{a}$ are replicated fermion fields and $S_{{\rm dis},\mu}$ is an elastic four fermion interaction term proportional to $\KK_{\mu}$. Upon integrating out the energies $e^{-l}\Lambda<|\varepsilon_{\pm}|/\hbar v<\Lambda$ in one-loop approximation and rescaling momenta $\mathbf{q}\rightarrow e^{l}\mathbf{q}$  and frequencies $\omega\rightarrow e^{lz}\omega$ with $z=1+ \sum_{\mu=0,x,y,z} \KK_{\mu}/2 \pi^2$, we find the flow equations 
\begin{eqnarray}
\partial_{l}\KK_{0} & = & -\KK_{0}+ \case{\KK_{0}}{\pi^2} 
  \Big(\KK_{0}+ \sum_{j=x,y,z} \KK_{j} \Big),\label{eq:dK0}\\
\partial_{l}\KK_{i} & = & -\KK_{i} - \frac{\KK_{i}}{3 \pi^2}
  \Big( \KK_{0} + 2 \KK_{i} - \sum_{j=x,y,z}\KK_{j} \Big),\label{eq:dKi}
\end{eqnarray}
with $i=x,y,z$. 

According to the flow equations \eqref{eq:dK0} and \eqref{eq:dKi} all weak disorder is irrelevant, consistent with the general expectations based on the scaling dimension of the disorder term. Although the flow equations are valid up to second order in the disorder strengths $\KK_{\mu}$ only, it is instructive to analyze in which cases they predict a transition to a strong-disorder phase. Here, an important observation is that disorder types not present initially will not be generated along the flow, so that we may gain a good understanding by considering different numbers of disorder components separately. 

(i) If there is scalar disorder only ($\KK_i=0$) the flow equation (\ref{eq:dK0}) reproduces the flow equation of Ref.\ \cite{Syzranov2015a}, which predicts a critical point at $\KK_{0}= \KK_{\rm c} \equiv \pi^{2}$. In contrast, the flow equations predict no disorder-induced transition for pure single-component vector disorder, such as the case of nonzero $\KK_x$ with $\KK_y=\KK_z=0$. 

(ii) For scalar disorder along with a single vector component $\KK_{i}$, the processes mentioned above are mutually enhanced: The presence of the scalar component $\KK_{0}$ makes the vector component $\KK_{i}$ even more irrelevant, while the presence of the vector component $\KK_{i}$ lowers the critical disorder strength for the scalar component. For pure vector disorder with two components, say $\KK_{x}$ and $\KK_{y}$, the relative difference $s=(\KK_x-\KK_y)/(\KK_x+\KK_y)$ satisfies the flow equation
$\partial_{l} s = - s (1 - s^2) (\KK_x+\KK_y)/3 \pi^2,
$
so that for strong disorder $s$ rapidly approaches zero. This motivates setting $\KK_x = \KK_y = \KK_{xy}$, which has the flow equation $\partial_{l}\KK_{xy}=-\KK_{xy}$. Without the second-order term, the third-order term determines the presence of a critical point. While we have not performed the corresponding two-loop calculations, numerical results below give evidence for a diffusive phase in this case.

(iii) For three or four disorder components, scalar disorder, if present, will eventually dominate the flow to a diffusive phase. For pure three-component vector disorder similar arguments suggest that the intermediate flow is towards the case $\KK_{x}=\KK_{y}=\KK_{z}\equiv \KK_{xyz}$, which has the flow equation $\partial_{l}\KK_{xyz}=-\KK_{xyz}+\KK_{xyz}^{2}/3 \pi^2$, with a critical disorder strength $\KK_{xyz}=3 \KK_{\rm c}$.

\paragraph*{Density of states.---}

\begin{figure}
\noindent \begin{centering}
\includegraphics{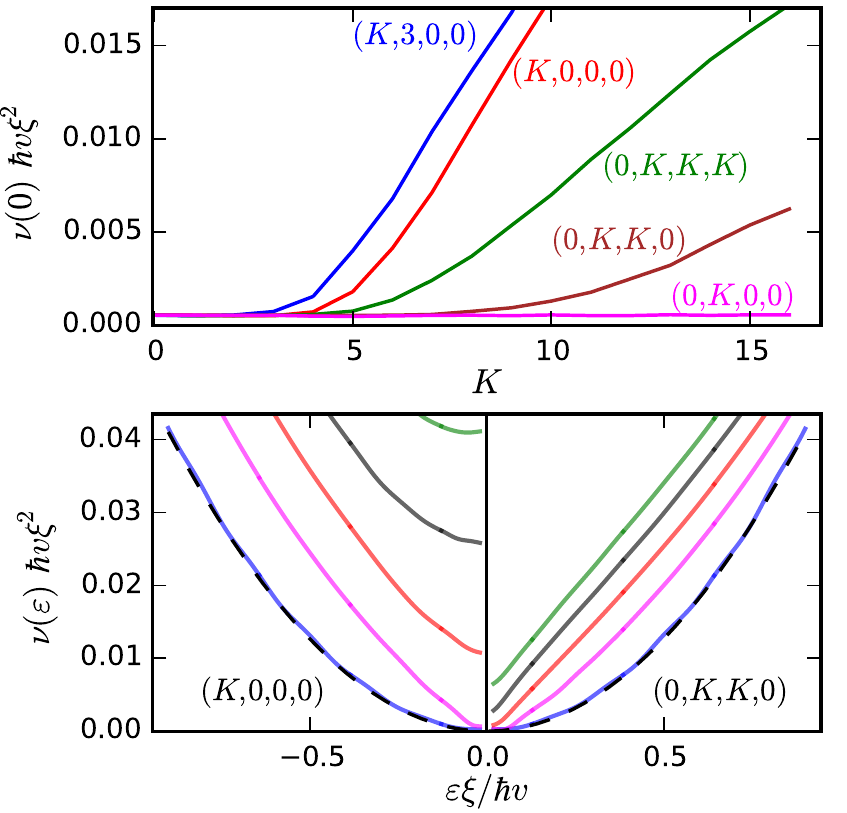}
\par\end{centering}

\protect\caption{\label{fig:KPM}(color online) Top panel: Density of states $\nu(0)$ at the nodal point and per Weyl node for various disorder types $(K_0,K_x,K_y,K_z)$ as a function of disorder strength $K$, as indicated in the figure. Bottom panel: $\nu(\varepsilon)$ versus energy $\varepsilon$ for scalar disorder (left) and for two-component vector disorder with $K_x=K_y=K_{xy}$ (right). The disorder strengths $K_0$ or $K_{xy}$ are $0,4,8,12,16$ (bottom to top). The density of states $\nu_0(\varepsilon) = \varepsilon^2/2 \pi^2 (\hbar v)^3$ of the clean Hamiltonian (\ref{eq:H0}) is shown dashed. 
The numerical calculation was performed for a cube geometry with linear size $L = 200 a$. The number of random vectors used for calculating the trace in the kernel polynomial method is $20$ \cite{Weisse2006}. The energy resolution at the nodal point is $\Delta\varepsilon\simeq0.024\,\hbar v/\xi$. The data represents an average over \textasciitilde{}10 disorder realizations. The finite offset value for $\nu(0)$ that is observed even for $K=0$ is a finite-size effect.}
\end{figure}

To complement the results of the scaling analysis we calculate the density of states $\nu(\varepsilon)$ in a disordered tight-binding model of a Weyl semimetal using the kernel polynomial method \cite{Weisse2006}. The density of states $\nu(0)$ at the nodal point serves as an order parameter for the semimetal-to-diffusive metal transition \cite{Fradkin1986,Fradkin1986a,Kobayashi2014}. We use a lattice version of the Hamiltonian (\ref{eq:H0}), 
\begin{equation}
H_{0,{\rm lattice}} = \frac{\hbar v}{a} \left(\sigma_{x}\sin ak_{x}+\sigma_{y}\sin ak_{y}-\sigma_{z}\cos ak_{z}\right),\label{eq:HL}
\end{equation}
where $a$ is the lattice constant. This Hamiltonian has eight Weyl points at crystal momenta $\mathbf{k}_{\eta_x,\eta_y,\eta_z} = (\pi/2 a) (1+\eta_x,1+\eta_y,\eta_z)$, with $\eta_{x,y,z} = \pm 1$. We add Gaussian-distributed disorder of the form \eqref{eq:U(r)}, with zero mean and with correlation function
\begin{equation}
\left\langle U_{\mu}(\mathbf{r})U_{\nu}(\mathbf{r}^{\prime})\right\rangle =\delta_{\mu\nu}K_{\mu}\frac{\left(\hbar v\right)^{2}}{(2 \pi)^{3/2} \xi^{2}}e^{-|\mathbf{r-r^{\prime}}|^{2}/2\xi^{2}}, \label{eq:correlated_dis}
\end{equation}
where $\xi$ is the disorder correlation length and $K_{\mu}$ the dimensionless disorder strength. (The dimensionless disorder strength $K_{\mu}$ is similar to the dimensionless disorder strength $\kappa_{\mu}$ in the scaling approach, but may differ quantitatively because of the different short-distance regularizations in the two approaches.) We choose $\xi = 5a$ to suppress the inter-node scattering rate, thus realizing effectively single-node physics, compatible with the scaling approach above and the transport calculations below \footnote{For this value of $\xi$, the ratio of inter-node and intra-node scattering rates is less than $10^{-19}$.}. Results for the density of states $\nu(\varepsilon)$, normalized to a single Weyl node and averaged over 10 disorder realizations, are shown in Fig.\ \ref{fig:KPM}, bottom panel. The density of states at the Weyl point $\nu(0)$ is shown in the top panel of the same figure. The $\nu(0)$ data qualitatively confirm the conclusions drawn from the flow equations. In particular, adding a single-component vector disorder to scalar potential disorder lowers the critical disorder strength; the critical disorder strength for pure vector disorder with $K_{x}=K_{y}=K_{z}=K_{xyz}$ is higher than for purely scalar disorder; no diffusive phase is seen for single-component pure vector disorder. The numerical results further indicate that vector disorder with two nonzero components $K_{x}=K_{y}=K_{xy}$ drives the system into a diffusive phase for $K_{xy}\apprge9$, corresponding to an instability deriving from higher-order terms in the flow equation that are not captured in our one-loop perturbative analysis.

The bottom panel of Fig. \ref{fig:KPM} compares the density of states $\nu(\varepsilon)$ around the Weyl point $\varepsilon=0$ for scalar disorder ($K_0$) and for pure two-component vector disorder ($K_x=K_y=K_{xy}$). The disorder-induced increase of the density of states at finite energy can be understood as the result of a renormalization of the Fermi velocity, consistent with the transport data that follow below. (An analytical assessment of velocity renormalization would require the inclusion of two-loop diagrams in the RG calculations, which is beyond the scope of this article.) Around the critical disorder strengths, we find a singularity in the density of states (smoothed by the finite resolution of the kernel polynomial method) that is much steeper for the two-component vector case than for the scalar case (compare, {\em e.g.} the data for $K_{0}=4$ and $K_{xy}=12$). This difference is in qualitative agreement with the prediction from a scaling Ansatz around criticality, $\nu(\varepsilon)\propto\left|\varepsilon\right|^{(3-z)/z}$ \cite{Kobayashi2014}, where $z=1+\sum_{\mu}K_{\mu}/2\pi^{2}$ evaluated at the critical disorder strength, is the dynamical critical exponent. Thus, the larger $\sum_{\mu}K_{\mu}$ at criticality, the sharper the singularity.

\paragraph*{Quantum transport.---} Calculations of quantum transport properties at the nodal point provide an alternative route towards the observation of the disorder-induced phase transition. We consider a Weyl semimetal attached to ideal source and drain leads, with dimensions $L$ and $W$ in the transport and transverse directions, respectively. In the pseudo-ballistic weak disorder phase, one expects the same transport characateristics as those of the clean Hamiltonian (\ref{eq:H0}): conductance $G \propto W^2/L^2$ and Fano factor $F = F_0=1/3 + 1/6 \ln 2 \approx 0.574$ \cite{Baireuther2014,Sbierski2014a}. (The Fano factor is the ratio of the shot noise power and the current, see, {\em e.g.}, Ref.\ \cite{blanter2000b}.) In contrast, in the diffusive regime $G$ is proportional to $W^2/L$, corresponding to a finite conductivity $\sigma$, whereas $F = 1/3$ \cite{beenakker1992}.

The transport calculations with a random vector potential closely follow our previous calculations for a Weyl semimetal with scalar disorder only \cite{Sbierski2014a}. We apply periodic or antiperiodic boundary conditions in the directions transverse to the current flow. The conductance $G$ (per Weyl node) and the Fano factor $F$ are calculated from the transmission matrix $t$ as $G= (e^2/h) \mathrm{tr} tt^{\dagger}$, $F=1-\mathrm{tr}(tt^{\dagger})^2/\mbox{tr}\, tt^{\dagger}$. Anticipating a scaling $G \propto W^2$ we normalize the conductance to a cube geometry, $G = W^2 G_{\rm cube}/L^2$. We take the sample width $W$ and the size $M$ of the transmission matrix large enough that our results for $G_{\rm cube}$ and $F$ do not depend on these, nor on the choice of the boundary conditions in the transverse direction. In contrast to the density of states calculation above, the fixed transport direction $(z)$ requires us to distinguish the longitudinal $(z)$ and transverse ($x$,$y$) components of the vector disorder. 

\begin{figure}
\noindent \begin{centering}
\includegraphics{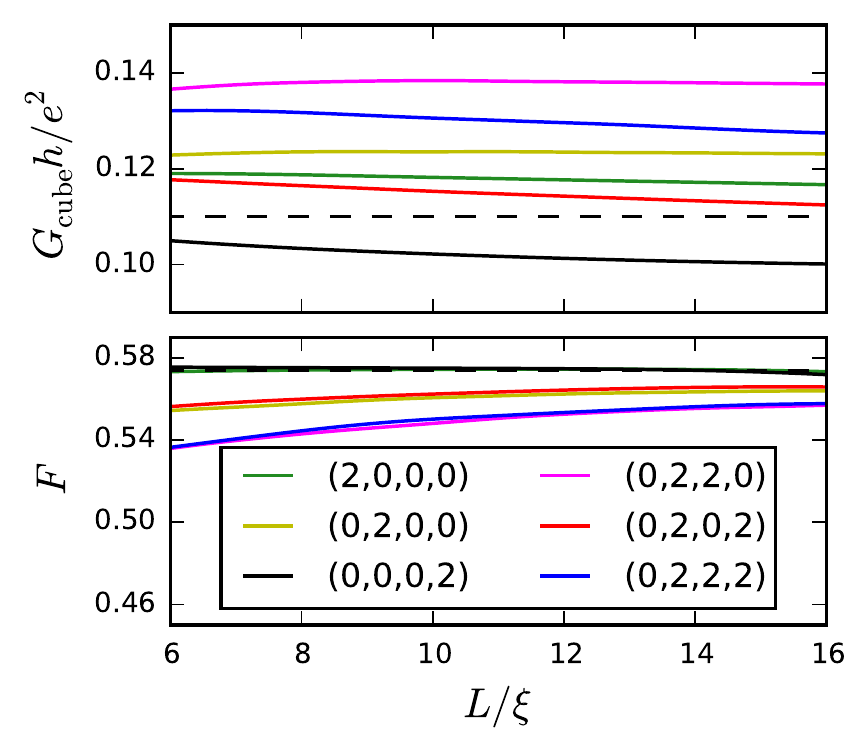}
\par\end{centering}

\protect\caption{{\small{}\label{fig:transport_weak_K}(color online) Cube conductance $G_{\rm cube}$ and Fano factor $F$ as a function of sample size $L$ for a single Weyl node with various types of weak disorder potentials in the pseudoballistic transport regime, with disorder strengths $(K_0,K_x,K_y,K_z)$ as indicated in the figure. Numerical data are from an average over at least 10 disorder realizations and over periodic and antiperiodic boundary conditions. The asymptotic values $G_{0,\mathrm{cube}}$ and $F_{0}$ for transport in a clean, isotropic Weyl node are denoted by dashed lines.}}
\end{figure}

Results for weak disorder are shown in Fig.\ \ref{fig:transport_weak_K}. We observe that $F\rightarrow F_{0}$ for large system size $L$ and that $G_{\rm cube}$ approaches a constant for large $L$, although our numerics indicates that the asymptotic value for $G_{\rm cube}$ may differ from the size-independent cube conductance of the clean limit $G_{0,{\rm cube}} = (1/2 \pi) \ln 2 \approx 0.11$ \cite{Baireuther2014,Sbierski2014a}. This difference is consistent with the possibility of an anisotropic disorder-induced renormalization of the Fermi velocity $v$ to smaller values. Indeed, without disorder, an anisotropic change of the velocities $v \to v_i$, $i=x,y,z$ was found to give $G_{\rm cube} = G_{0,{\rm cube}} v_z^2/v_x v_y$, consistent, {\em e.g.}, with a decrease (increase) of $G_{\rm cube}$ for single-component vector disorder with $K_z \neq 0$ ($K_x \neq 0$) below (above) $G_{0,{\rm cube}}$, wheraes the Fano factor $F$ remained unaffected \cite{Trescher2015}.

\begin{figure}
\noindent \begin{centering}
\includegraphics{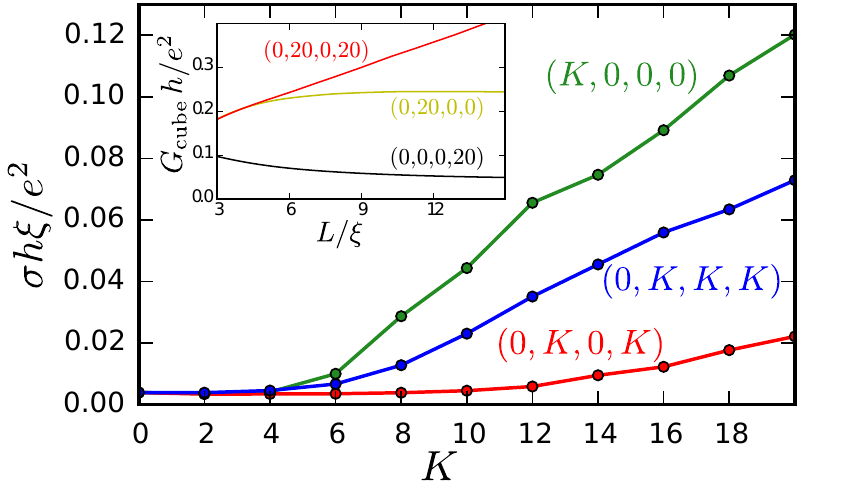}
\par\end{centering}

\protect\caption{\label{fig:transport_strong_K}(color online) Conductivity $\sigma$ as a function of disorder strength $K$ for a single Weyl node with various types of disorder potentials $(K_0,K_x,K_y,K_z)$ as indicated in the figure. The inset depicts cube conductance $G_\mathrm{cube}$ as a function of sample size $L$. The data represent an average over at least 10 disorder realizations and over periodic and antiperiodic boundary conditions. The finite offset value for $\sigma$ that is observed even for $K=0$ is a finite size effect.}
\end{figure}

For large disorder strengths, we find that the pseudoballistic transport characteristics are preserved for all disorder strength for single-component pure vector disorder (see Fig.\ \ref{fig:transport_strong_K}, inset). All other disorder types show a diffusive scaling $G = \sigma W^2/L$, $\sigma$ being the bulk conductivity, and $F\rightarrow 1/3$ above a critical disorder strength. The main panel in Fig.\ \ref{fig:transport_strong_K} shows our numerical results for the conductivity, obtained from the relation $\sigma^{-1}=W^{2}\partial_{L}G^{-1}(L)$. The critical disorder strengths obtained from the conductivity data are in good quantitative agreement with those obtained from the density of states calculations, see Fig.\ \ref{fig:KPM}, top panel.

\paragraph*{Conclusion.---}
We have shown that the inclusion of a random vector potential leads to a rich phase diagram for a disordered Weyl node. Whereas weak disorder is always irrelevant, the scaling equations and numerical data presented here indicate that for a random vector potential there is a disorder-induced transition to the diffusive phase only if the number of statistically independent components of the random vector potential is two or more. Our transport data indicates that the nature of the strong-disorder phase is a diffusive metal, regardless of the disorder types present.

Although the critical points observed in this study all separate a pseudoballistic from a diffusive phase, it is likely that they belong to different universality classes. This is plausible since the scalar disorder induced critical point is time-reversal symmetric while a more generic fixed point at a finite amout of vector disorder breaks this fundamental symmetry. Further evidence was given by the analytical and numerical assessment of the dynamical critical exponent $z$. The fact that the flow equations give correlation length exponents $\nu=1$ for all disorder types must be regarded as an artifact of the one-loop approximation \cite{Syzranov2016}. Further work is required to pin down these critical exponents.

Another interesting aspect of the vector disorder induced phase transitions pertains to rare region effects. For potential disorder, it has been argued \cite{Nandkishore2014,Pixley2016} that a nonzero (but exponentially small) density of states at zero energy persists for subcritical disorder strengths, caused by states trapped in exceptionally strong potential fluctuations. While the consequences of this claim are still under debate, see, {\em e.g.} Ref.\ \cite{Pixley2016}, we point out that the same type of argument cannot be staightforwardly carried over to a pure vector potential, since even a locally strong vector potential does not trivially lead to a finite density of states at the nodal point. It is an interesting question, whether a rare-region analysis confirms the qualitative differences between single-component and multicomponent vector disorder.  

We close by pointing out differences and similarities with the case of a two-dimensional Dirac cone,
\begin{equation}
\tilde{H}=\hbar v\left(\sigma_{x}k_{x}+\sigma_{y}k_{y}\right)\label{eq:2dDirac}
\end{equation}
with disorder as in Eq. \eqref{eq:U(r)}, a problem that has received enormous interest in the theoretical literature, see Ref.\ \cite{Ludwig1994} for an early study and \cite{Evers2008} for a review. By power counting, in two dimensions all disorder is marginal at tree level. In the clean case, Eq.\ \eqref{eq:2dDirac} is a critical theory at a topological transition tuned by a mass term $M\sigma_{z}$ that changes the Chern number by one. Disorder of potential- or ``mass''-type, with strengths $K_{0}$ and $K_{z}$, is marginally relevant and irrelevant, respectively. Any combination of vector disorders ($\propto \sigma_x$ or $\sigma_y$) is a renormalization-group fixed-point, that moreover affords an exact solution with multifractal wavefunctions. Accordingly, the density of states scaling $\nu(\varepsilon)\propto\left|\varepsilon\right|^{(2-z)/z}$ is valid for any vector disorder and not just around a critical disorder strength as in the Weyl case. Moreover, by applying a pseudo-gauge transformation in the quantum transport setup at zero energy, it can be shown that pure vector disorder does not affect any transport properties of \eqref{eq:2dDirac} \cite{Schuessler2009}, quite in contrast to our findings for the three-dimensional Weyl case. Another difference between the two-dimensional and three-dimensional cases is that if two of the three disorder couplings $K_{0},\, K_{z},\, K_{\perp}\equiv K_{x}+K_{y}$ are nonzero initially, in the two-dimensional case the remaining component gets generated along the flow \cite{Schuessler2009}. In contrast, the flow equation for the three-dimensional Weyl case does not generate new coupling constants.

\begin{acknowledgments}
\paragraph*{Acknowledgments.---}

Numerical computations were done on the HPC cluster of Fachbereich Physik at FU Berlin. Financial support was granted by the Helmholtz Virtual Institute ``New states of matter and their excitations'' and by the Emmy Noether program (KA 3360/2-1) and the CRC/Transregio 183 (Project A02) of the Deutsche Forschungsgemeinschaft.
\end{acknowledgments}

\bibliographystyle{apsrev4-1}
\bibliography{library}

\end{document}